# Phase boundary anisotropy and its effects on the maze-to-lamellar transition in a directionally solidified Al-Al$_2$Cu eutectic

U. Hecht[1a], J. Eiken[a], S. Akamatsu[b], S. Bottin-Rousseau[b]

[a]Access e.V., Intzestr. 5, 52072 Aachen, Germany

[b]Institut des Nanosciences de Paris, CNRS UMR 7588, Sorbonne Université, 4 place Jussieu, 75252 Paris Cedex 05, France

**Abstract:** Solid-solid phase boundary anisotropy is a key factor controlling the selection and evolution of non-faceted eutectic patterns during directional solidification. This is most remarkably observed during the so-called maze-to-lamellar transition. By using serial sectioning, we followed the spatio-temporal evolution of a maze pattern over long times in a large Al-Al$_2$Cu eutectic grain with known crystal orientation of the Al and Al$_2$Cu phases, hence known crystal orientation relationship (OR). The corresponding phase boundary energy anisotropy (γ-plot) was also known, as being previously estimated from molecular-dynamics computations. The experimental observations reveal the time-scale of the maze-to-lamellar transition and shed light on the processes involved in the gradual alignment of the phase boundaries to one distinct energy minimum which nearly corresponds to one distinct plane from the family $\{120\}^{Al} \parallel \{110\}^{Al2Cu}$. This particular plane is selected due to a crystallographic bias induced by a small disorientation of the crystals relative to the perfect OR. The symmetry of the OR is thus slightly broken, which promotes lamellar alignment. Finally, the maze-to-lamellar transition leaves behind a network of fault lines inherited from the phase boundary alignment process. In the maze pattern, the fault lines align along the corners of the Wulff shape, thus allowing us to propose a link between the pattern defects and missing orientations in the Wulff shape.

---

[1] Corresponding author: u.hecht@access-technology.de

**1. Introduction**

Coupled growth of eutectic alloys is one of the well-known solidification processes that lead to pattern formation phenomena that are governed by diffusion and capillarity [1]. Eutectic growth has been extensively studied theoretically, experimentally and numerically in model systems that display isotropic properties of the involved interfaces, both solid-liquid and solid-solid. The isotropic-interface assumption is essentially valid as concerns the solid-liquid interfaces in the so-called "regular" eutectics. In contrast, the solid-solid phase boundaries in the eutectic most often present a non-negligible free-energy anisotropy, which depends on the relative orientation of the solid crystals in a given eutectic grain. This anisotropy can have a major impact on the morphological stability of coupled-growth patterns and their dynamic response to changing growth conditions [2-6].

Our current knowledge on phase boundary anisotropy in eutectic alloys is rather limited despite the vast literature on characteristic crystal orientation relationships (ORs) in eutectic alloys e.g. [7-10]. A particular OR is defined by common lattice planes of the two crystals, and by a common direction in that plane. It is current practice to label an OR by the Miller indices of the common planes and directions. In many cases, the common planes belong to families of lattice planes that are relatively dense, and present rather low misfit parameters. It is therefore reasonable to assume that the common planes correspond to a low free energy configuration. However, quantitative information on the full phase boundary anisotropy for a given OR, meaning the dependence of the free energy of the phase boundary as function of its location relative to a selected coordinate system, is generally unknown.

One exception is the Al-Al$_2$Cu system that is widely used as a model system to study eutectic pattern formation during directional solidification. The phases involved in coupled eutectic growth are the face centered cubic crystal of the Al solid solution (space group no. 225) and the ordered intermetallic compound Al$_2$Cu with a tetragonal crystal lattice (space group no. 194) as described in [11]. For this eutectic and specifically for two distinct ORs in this system the phase boundary energy landscape has been computed at T=50K using molecular dynamics [12]. Figure 1 illustrates the data for the so-called "Alpha-4" OR of interest in the present work, which is found when the Al and Al$_2$Cu crystalline phases in a eutectic grain are aligned relative to one another as in fig. 1(a) with the common planes being $\{130\}^{Al}$ and $\{100\}^{Al2Cu}$ and the common direction $[001]^{Al}$ ∥ $[001]^{Al2Cu}$. As can be seen from the pole figures of fig. 1(b), the common planes of this family appear at distinct locations, corresponding to azimuthal angles of 0°, 90°, 180° and 270°. Interestingly, the Alpha-4 OR offers another 4-fold family of common planes $\{120\}^{Al}$ ∥ $\{110\}^{Al2Cu}$ with the same common direction as above (Fig.1c). The common planes of this family appear at azimuthal angles of 45°, 135°, 225°and 270°. The calculated 2D γ-plot for this OR [12] is displayed in fig. 1(d). It is obviously anisotropic, meaning that the phase boundary energy depends on its location relative to an orthogonal coordinate system chosen to be the unit cell of the Al₂Cu crystal. The γ-plot presents eight shallow energy minima labelled A1 through A4 and D1 through D4. Eutectic phase boundaries may thus prefer to align to any of these minima in what is called a maze pattern. The γ-plot displays a central symmetry but no mirror symmetry and therefore any application to experimental situations requires a careful alignment to the given crystal orientation. Note that the MD computations [12] for this crystal alignment and OR yield minima slightly off-set from the common plane locations (minima at irrational planes).

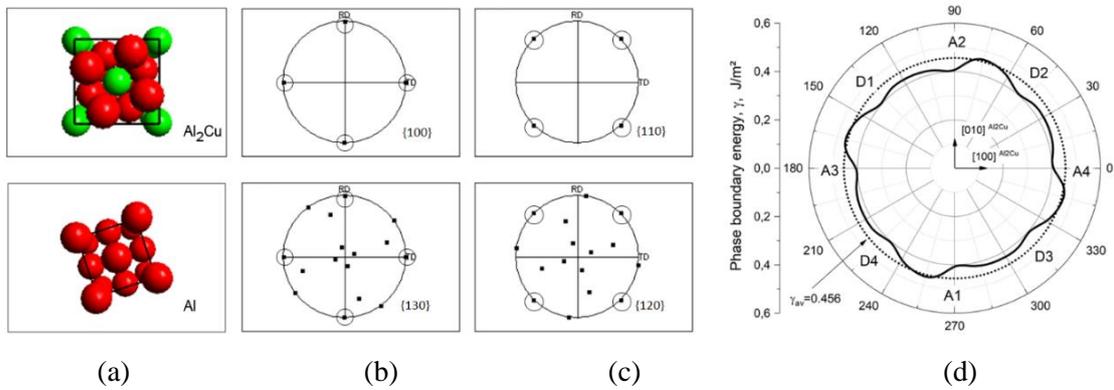

**Fig. 1:** Crystal orientation relationship "Alpha-4": **(a)** crystal mimic, **(b)** pole figures $\{100\}^{Al2Cu}$ and $\{130\}^{Al}$, **(c)** pole figures $\{110\}^{Al2Cu}$ and $\{120\}^{Al}$, **(d)** 2D γ-plot of the phase boundary energy, also provided as data set in the supplementary material. The common direction is $[001]^{Al2Cu} \parallel [001]^{Al}$ being parallel to the growth direction.

On this background the objective of this work is to study the lamellar pattern evolution during early stages of directional solidification in a eutectic grain with the Alpha-4 OR with main emphasis on the maze-to-lamellar transition. We will ask and experimentally investigate the following questions:

(1) Why are lamellar grains obtained at all, given the distinct locations of 2x4 energy minima in the γ-plot? What is driving the selection of one out of several alignment options?

(2) What are the pattern adjustment mechanisms and the time-scale of the maze-to-lamellar transition?

(3) What defect structures develop along with the maze-to-lamellar transition?

(4) Can maze patterns be regarded as the fingerprint of phase boundary anisotropy and if yes can their analysis provide sufficient information about the γ-plot even if this was unknown?

The study will be based on the 2D γ-plot, because directional solidification experiments reveal that for this OR the $[001]^{Al} \parallel [001]^{Al2Cu}$ common directions are almost always parallel to the main solidification axis **z**. The common planes are therefore parallel to the direction of growth, and the phase boundaries do not present a substantial inclination with respect to the **z** axis. This has been verified experimentally by focused ion beam cutting (FIB), without being shown here.

Maze patterns have been investigated in isotropic and anisotropic eutectics before [13-15], and we briefly recall the already known characteristics of the maze-to-lamellar transition: Perrut et al. [13] showed that for the isotropic case a maze pattern will persist for indefinitely long times unless an external bias introduces a symmetry break and hence a preferred direction for lamellar alignment. The results are based on *in-situ* experimental observations in an organic eutectic alloy biased by tilting the isotherms in the solidification set-up, e.g. by imposing a macroscopic tilt or curvature to the solidification front. Ghosh et al. [14, 15] further extended the work as to include phase boundary anisotropy: a phase field study revealed that an imposed 4-fold anisotropy of the solid-solid phase boundary energy leads to a maze with rectangular features of the pattern, including lamellae with sharp bends or corners as well as sharp lamellae ends. The mazes did not evolve into a fully ordered lamellar pattern, irrespective of the probed anisotropy strength. At contrary, a 2-fold anisotropy induces a favorable orientation of the lamellar interfaces and a fast alignment of the phase boundaries towards the energy minimum in early stages; the kinetics is however time dependent and slows down quickly. To our knowledge no experimental research has been published on the time evolution of anisotropic mazes and the emerging lamellar order, though mazes as such have been observed and described qualitatively before [16].

## 2. The maze-to-lamellar transition: experimental procedure and analysis methods

### *2.1 Experimental procedure*

Unidirectional solidification in a Bridgman furnace with liquid metal cooling was used to process a bulk sample (diam. 8 mm, length 220 mm) under a temperature gradient of G=27 K/mm and an imposed pulling velocity of v=2 µm/s. The alloy composition was Al-17.5 at.%Cu-1 at.% Ag. As common in Bridgman experiments the sample was molten directionally to about 2/3$^{rd}$ of its length, followed by a holding stage of 10 minutes for thermal equilibration. During equilibration a thin $Al_2Cu$-layer developed at the fusion front [16] due to thermo-diffusion of Cu in the liquid phase [17] under the applied temperature gradient. From this "seed" eutectic coupled growth evolved (i) during the initial transient and further (ii) in steady state growth conditions. The experiment was monitored *in-situ* using the ultrasonic pulse-echo method [18] to measure the position of the solidification interface as it advances in time. This allowed estimating the local growth velocity in the initial transient stage and allocating the correct time to any position in the sample, e.g. to any transverse section plane.

Metallographic section planes were prepared from the region of the initial transient by serial sectioning, e.g. by successively grinding and polishing a new section plane for microstructure analysis. In total 12 section planes were prepared, labelled E1 through E12, being aligned perpendicular to the direction of solidification. The normal direction to the polished surface pointed in direction opposite to the direction of solidification. A reference plane for correct alignment of each section plane relative to one another was created before by milling in longitudinal direction. Fig. 2a gives an overview of the position and solidification time for each of the section planes.

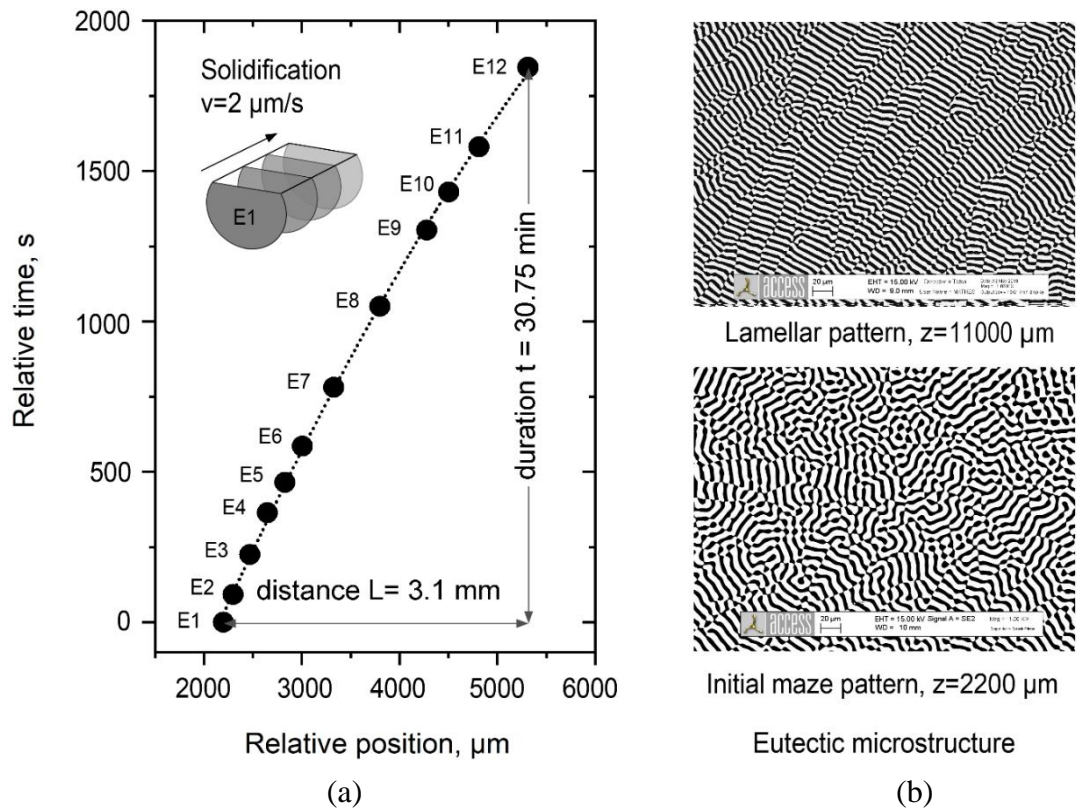

(a)            (b)

**Fig. 2: (a)** Location of the experimentally selected serial sectioning planes E1 through E12 along the sample length relative to the onset of directional solidification. The investigated length spans over a distance of 3.1 mm corresponding to a duration of 30.75 minutes in the initital transient under the applied pulling velocity of 2 µm/s; **(b)** eutectic microstructure in the initial maze at E1 (z=2.2 mm) and in the fully developed lamellar grain at a reference position z=11 mm. The lamellar spacing λ at the reference position ranges around 7 µm. Note that the lamellar phase boundaries at z=11000 µm are still not fully aligned to a single orientation.

The position E1 corresponds to z=2.2 mm measured from the fusion front. The series of sections encompasses a solidification distance of L=3.1 mm corresponding to a duration of t=30.75 minutes. Fig.2b displays an overview of the microstructure from section E1 in the region selected for further analysis. It shows the characteristic features of an anisotropic maze pattern. The fully developed lamellar structure in the region of interest after 11 mm of growth is equally shown in fig. 2b. Note that at this position the lamellar phase boundaries at are still not fully aligned to a single orientation.

*2.2 Microstructure analysis*

Each section was metallographically prepared using standard preparation techniques, including grinding and polishing. One after the other the sections were observed in an SEM type Zeiss Gemini 1550 SEM after careful alignment and navigation to the target position containing the region of interest. Backscatter electron images (BSE) were taken each time. The last section E12 was specially prepared for electron backscatter diffraction measurements (EBSD) by ion milling using argon ions within a GATAN Model 682 ion milling device. The Oxford INCA Crystal software was used to record and evaluate the Kikuchi patterns acquired from the region of interest with a HKL Nordlys detector. The EBSD mapping results were used to determine the crystal orientation and orientation relationship, by searching for superposing poles. The fully developed lamellar structure obtained after 11 mm of growth was analyzed for reference by EBSD and image analysis.Digital image analysis was used to measure the local phase boundary orientations and provide angular distribution histograms and texture maps for the acquired BSE-images. Two analysis algorithms were used, (i) *Gabor filter* banks implemented in Python for the texture analysis and (ii) a *Structure Tensor* implemented in MATLAB. Both methods are well described in literature [19-21]. Both methods also provide a quality index, which is low in regions with corners, lamellar faults and other. The quality index images provide a good view of the defect distribution. All processed images were 324 x 324 pixel in size corresponding to a region of interest of 241 x 241 µm². The resolution was thus 0.744 µm / pixel. Selected micrographs along with texture maps and angular distribution histograms are displayed in fig. 5.

**3. The maze-to-lamellar transition: experimental results and discussion**

*3.1 Crystal orientation and OR determined by electron backscatter diffraction (EBSD)*

The crystal orientation and orientation relationship between Al and Al$_2$Cu in the region of interest was determined from EBSD measurements in section E12. Fig. 3 displays the EBSD results based on pole figures of the relevant families of planes along with the crystal mimic. The data show that the Al$_2$Cu-phase is aligned such that the $[00\bar{1}]^{Al2Cu}$ axis is nearly but not exactly parallel to the sample normal, in fact the Al$_2$Cu crystal texture reads $(10\bar{8})\langle 010\rangle^{Al2Cu}$. The Al-phase is aligned with the $[100]^{Al}$ parallel to the sample normal. Thus the two directions are slightly misaligned relative to one another, the disorientation angle being 2.5°. Nonetheless, the crystalline orientation in the region of interest is close to the Alpha-4 OR, since both families of common planes, $\{130\}^{Al} \parallel \{100\}^{Al2Cu}$ and $\{120\}^{Al} \parallel \{110\}^{Al2Cu}$ are present. The coordinates of the respective poles are listed in Table 1.

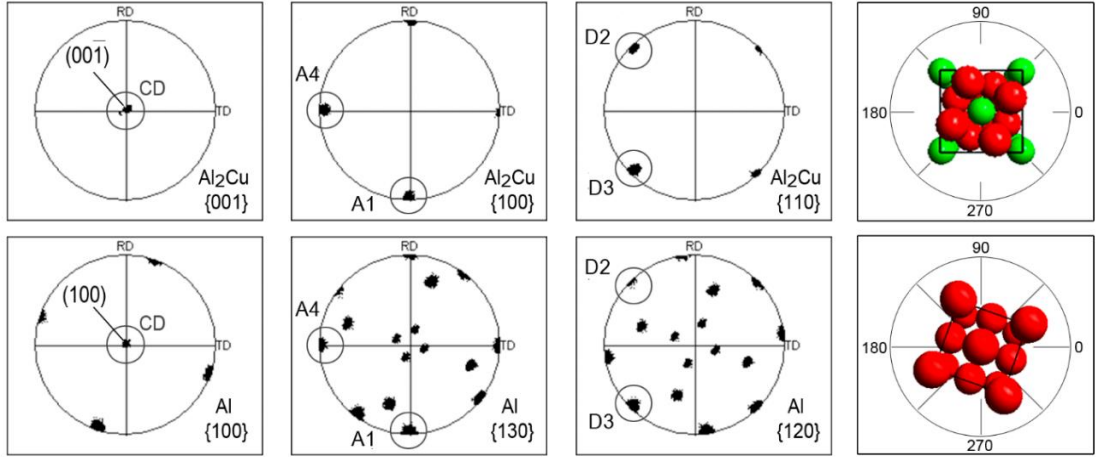

**Fig. 3**: Crystal orientation in the region of interest, measured by EBSD in section E12: shown are the relevant pole figures with the overlapping poles labelled as in fig. 4. The situation corresponds closely though not exactly to Al₂Cu unit cell axes being aligned as follows: $[00\bar{1}]^{Al2Cu} \parallel ND$; $[010]^{Al2Cu} \parallel RD$ and $[\bar{1}00]^{Al2Cu} \parallel TD$, such that the 2D-γ-plot applies.

Table 1: Overview of the EBSD results given as the median value from all individual data points in a pole cluster; the azimuthal angle is read from TD = 0°.

| OR – Alpha4, Common planes type A | | | | OR – Alpha4, Common planes type D | | | |
|---|---|---|---|---|---|---|---|
| Label | Planes | Azimuth, ° | Zenith, ° | Label | Poles | Azimuth, ° | Zenith, ° |
| A4 –Al₂Cu | $(100)^{Al2Cu}$ | 179.1 | 87.7 | D2 –Al₂Cu | $(110)^{Al2Cu}$ | 134.1 | 89.1 |
| A4 –Al | $(031)^{Al}$ | 179.8 | 90.0 | D2 –Al | $(02\bar{1})^{Al}$ | 134.8 | -89,4 |
| A4 | Bias | 0.7 | 2.3 | D2 | Bias | 0.7 | 2.5 |
| A1–Al₂Cu | $(0\bar{1}0)^{Al2Cu}$ | 269.1 | 88.9 | D3 –Al₂Cu | $(1\bar{1}0)^{Al2Cu}$ | 224.1 | 87.6 |
| A1–Al | $(0\bar{1}3)^{Al}$ | 269.7 | 87.7 | D3 –Al | $(012)^{Al}$ | 224.7 | 88.4 |
| A1 | Bias | 0.6 | 1.2 | D3 | Bias | 0.6 | 0.8 |

When carefully looking at the overlapping pole coordinates of the common crystal planes listed in Table 1 one observes a small bias corresponding to the disorientation between the common directions $[00\bar{1}]^{Al2Cu}$ and $[100]^{Al}$: the common crystal planes (poles) do not overlap equally well. The departure from the perfect crystallographic OR can be expressed as the difference between the coordinates of superposing poles in terms of their azimuthal and zenith angle. These differences are called "bias", since they break the symmetry of the OR. The smallest disorientation is found for the common pole D3, corresponding to the common plane $(1\bar{1}0)^{Al2Cu}$ ∥ $(012)^{Al}$ of the OR. Small crystallographic biases as above are likely to occur in real experimental conditions, but have not been probed by simulations as yet. In section 3.2 we will show that they impact on the alignment of phase boundaries during the maze-to-lamellar transition process. In fact we will conclude that the lamellar alignment is driven by this bias since indeed the Al-Al$_2$Cu phase boundaries gradually align to the common plane D3.

For the above experimental configuration the crystal mimic is just upside-down compared to fig. 1 and the γ-plot must be aligned to the unit cell axis of the Al$_2$Cu phase. The correctly aligned γ-plot is displayed in Fig. 4 along with the stiffness plot and the Wulff-plot including the Cahn-Hoffmann $\xi$-vector plot [22-24] to outline metastable and forbidden regions.

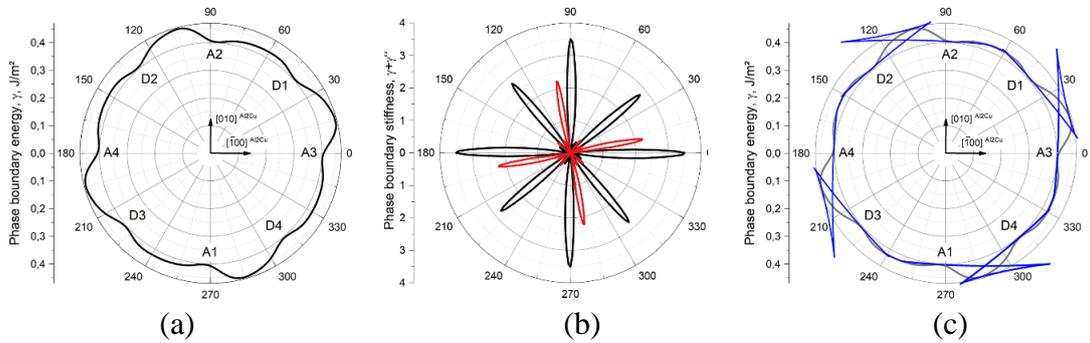

**Fig. 4:** Phase boundary energy landscape for the experimentally observed crystal orientation and OR including **(a)** the γ-plot, **(b)** the stiffness plot with negative stiffness values shown in red and **(c)** the Wulff-plot and the capillary $\xi$-vector plot showing "ears" associated with high negative stiffness values.

*3.2 Maze-to-lamellar transition as a fingerprint of anisotropic phase boundary energy*

The gradual alignment of phase boundaries during the maze-to-lamellar transition was analyzed using the *Gabor filter*, which operates like a bandpass filter with a series of differently oriented filter banks to reconstruct the local angular orientation from the response of the filter wavelets to the local gradients in the image. The *Gabor filter* was applied with an angular step of 2.5°, corresponding to 72 orientation classes from 0 to 1π. Other filter parameters were adjusted after having computed the dominant and direction-independent frequency by a Fast-Fourier-Transformation (FFT). The resulting texture maps and the associated quality index are presented in fig. 5 for selected section planes. The color coded maps represent the angle of the normal to the phase boundaries measured in clockwise direction relative to a horizontal line (TD in fig.3) thus encompassing the angular region from 0 to 180°. Fig. 5 also contains angular distribution histograms for each section plane which provide the relative frequency of the phase boundary orientation in the angular region from 360° to 180°. These histograms contain valid orientations only which have been determined using the *Structure Tensor* implemented in MATLAB. Taken together both methods show that the pattern is slowly evolving from a maze-like to a lamellar morphology. This is accomplished by aligning all phase boundaries to a rather narrow angular peak around 310 to 320°. The normal to the phase boundaries thus points to an azimuthal angle of 220 to 230° which in fact corresponds to the EBSD pole and the local energy minimum D3 (compare fig. 3 and table 1). All other phase boundary locations including those at the minima A4 //A3 and A1//A2 are slowly outgrown. This is conveniently seen from the texture maps: the red population with vertical phase boundaries and normal directions pointing to A4 // A3 disappear in time, despite of the fact that the corresponding local energy minimum is slightly lower than for D3 // D1, i.e. the stiffness is higher (compare fig.4).

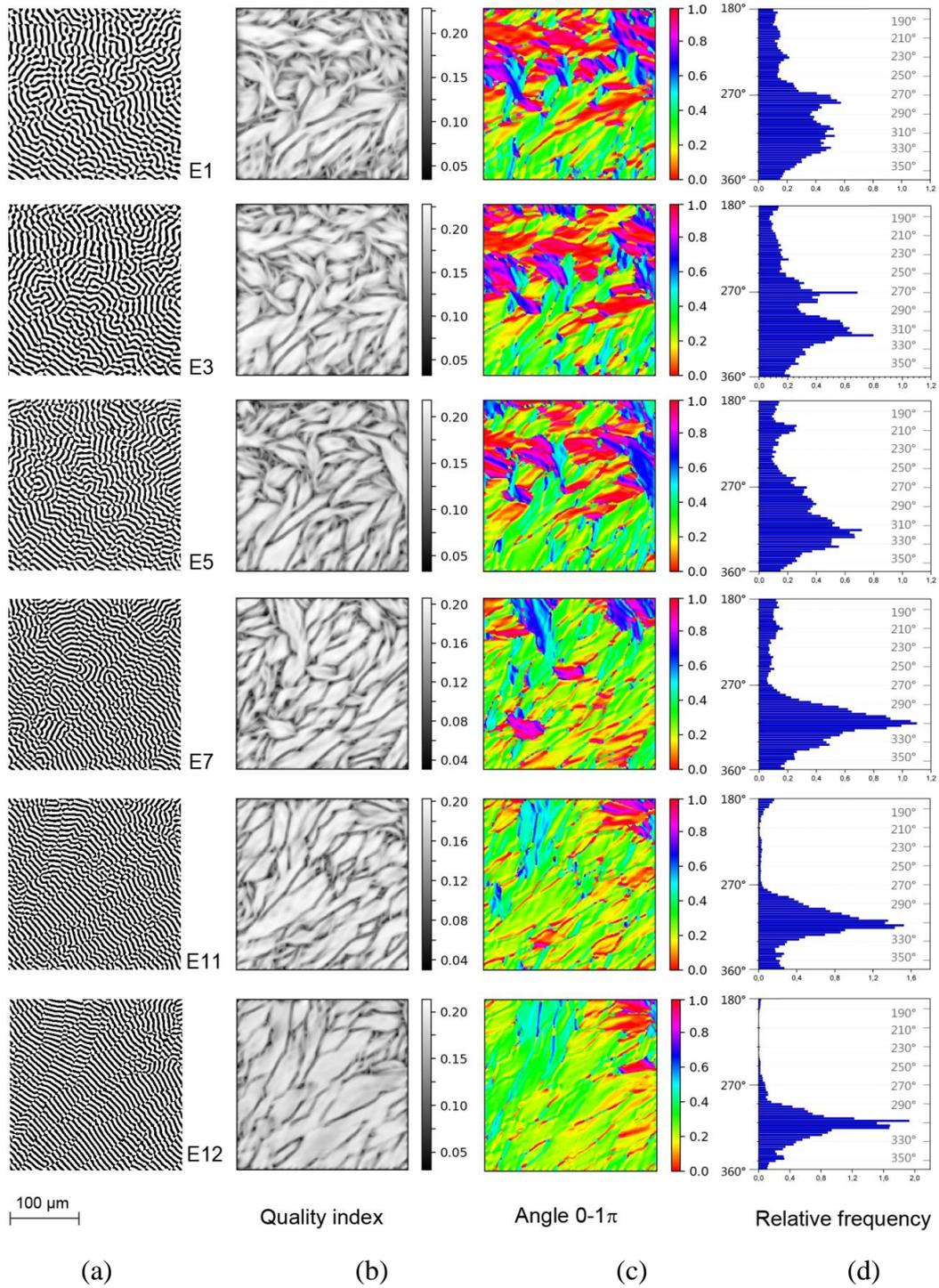

**Fig. 5:** Evolution of phase boundary orientation in selected section planes E1, E3, E5, E7, E11 and E12: **(a)** BSE-images in the region of interest; **(b)** *Gabor filter* quality index; **(c)** texture maps showing the angle of the normal to the phase boundaries from $0\pi=0°$ to $1\pi=180°$ and **(d)** angular distribution histograms showing the relative frequency of the phase boundary orientation as determined with the *Structure Tensor* analysis and plotted for the angular region from 360° to 180°. The angular resolution equals 2.5°.

The overall dynamics of the alignment is highlighted in fig. 6 taking into account all investigated serial sectioning planes: from the relative frequencies of the angular distribution histograms (fig.6a) the standard deviation of the frequencies was chosen as an integral measure of the degree of alignment. If phase boundary alignment is poor the relative frequencies for the distinct angle bins will not differ substantially and hence the standard deviation of the measured frequencies will be low. At contrary, pronounced peaks will lead to a higher value of the standard deviation. Fig. 6b shows the results plotted against the position corresponding to the successive section planes (compare fig. 2).

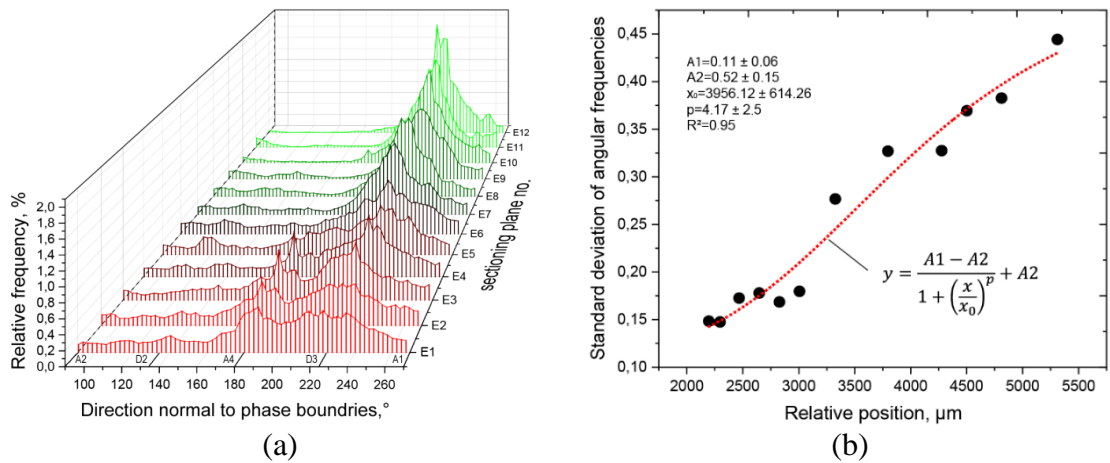

**Fig. 6:** Overview of the phase boundary alignment during the maze-to-lamellar transition for all serial sectioning planes expressed as relative frequency of the direction normal to the phase boundaries **(a)** and as standard deviation of the relative frequencies in time **(b).**

The data were fitted with an S-curve behavior to guide the eye and point out that the alignment process is slower in the beginning and also towards the end. One reason for the slow progress in early stages is most likely the presence of phase boundaries aligned to both A-type and D-type minima, often entangled in nodular core structures with characteristic angles. Once the A-type phase boundary populations disappears at about E6 the pattern reorganizes more easily. The process as such remains rather slow and is not finished at E12. The number of section planes is however not sufficient to finally conclude whether the transition kinetics follows a two-stage process or a smooth sigmoidal behavior. Future work should explore this aspect in more detail.

It is noteworthy that phase boundaries aligned to the A1//A2 and to D2//D4 minima are present in early maze patterns but their relative frequency is low, likely due to the given bias in the region of interest (see table 1). The question thus arises, whether experimental maze patterns contain sufficient information to serve as a fingerprint of the anisotropic phase boundary energy, e.g. whether they can be used to construct or validate the phase boundary energy landscape of the system at case. The limitation given by the bias can be overcome, knowing from EBSD measurements that the A-type and D-type families of common planes display a 4-fold symmetry, each. This means that the experimentally measured angular distribution of phase boundary orientations can be assigned to each quadrant of the polar plot. The results are shown in fig. 7 for the serial sectioning planes E1, E5 and E11. To ease the discussion the γ-plot from MD [12] has been added as well as the Wulff shape and the phase boundary stiffness plot to E11. For these plots the *Structure Tensor* analysis was run with a high angular resolution of 0.5° and again only valid orientations are plotted.

The polar plots show that the early maze pattern (fig. 7a) displays phase boundaries at virtually every angle, while the frequency shows distinct maxima and minima. During the maze to lamellar transition some angular regions disappear from the frequency plot (fig.7b) and the distribution sharpens until finally settling to a frequency maximum around the energy minima type D (here D3) as marked in fig. 7c.

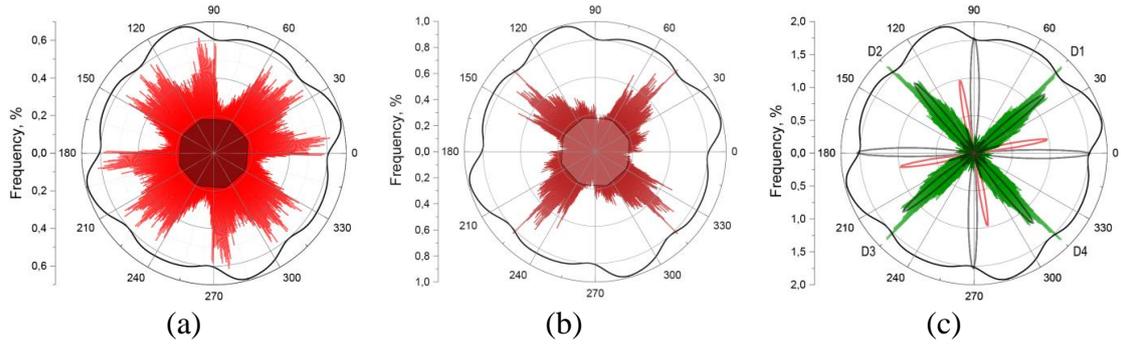

**Fig. 7**: Polar plots of the experimentally measured angular distribution of phase boundary orientations in section E1 **(a)**, E5 **(b)** and E11 **(c)** repeated in each quadrant based on the symmetry of the OR known from EBSD measurements. For convenience the γ-plot from MD computations [12] is included. The phase boundary stiffness plot is inserted in (c) while the Wulff shape is inserted in the center part of (a) and (b), respectively. Please note that by construction the 4 quadrants are identical.

From **fig. 7** the following features of the angular frequency plots in polar coordinates are outlined:

(i) The frequency of phase boundary orientations in the early maze structure (fig.7a) shows 8 maxima at locations which roughly point to the energy minima of the system; the four maxima around the azimuthal angles of 0, 90, 180 and 270° (type A) are sharper than the other four (type D). The type D maxima are spread over an angular range of about 30° around the azimuthal angles of 45°, 135°, 225° and 315°, respectively. It is obvious that the A-type frequency maxima are not symmetric around the A-type minima of the phase boundary energy plot, but slightly more distributed towards higher azimuthal angles. Upon closer inspection it turns out that the distribution is "skewed" towards the nearest-by corners of the Wulff shape (see central insert in fig.7a and fig.7b). This is likely also the reason why the D-type frequency maxima show three instead of one peak, the outer two peaks being close to the adjacent corners of the Wulff shape.

(ii) The frequency of phase boundary orientations in the early maze structure (fig.7a) shows 4 minima around the azimuthal angles of 20, 110, 200 and 290°, more specifically in regions corresponding to the corners with forbidden orientations (compare fig. 4c). However the distribution frequency is not zero at corners and phase boundaries running along the metastable directions of the *ξ*-vector plot are indeed observed in the micrographs. Noteworthy are the deep and wide frequency minima in the azimuthal angle range from 60° to 90° as well as the symmetric locations 150-180°, 240-270°and 330-360°. These minima were not expected but are thought to relate to the skewed distribution of phase boundaries towards the nearest-by corners of the Wulff shape (see central insert in fig.7a and fig.7b). Simulations would be requested to substantiate this observation, e.g. phase field simulations running on high power computational platforms.

(iii) Finally, the polar plots displayed in fig.7b and fig. 7c show that the angular distribution of phase boundaries sharpens as the pattern evolves from the initial maze towards a lamellar alignment. In the present case this goes along with the selection of the minimum D3 (and the theoretically symmetric D-type minima). The normal to the phase boundaries is mainly aligned to the D-type stiffness maxima in section E11 (fig. 7c) and additional sharp peaks are detected just aside, which point to the position of the $\{120\}^{Al} \parallel \{110\}^{Al2Cu}$ common plane location. Note that the directions of the D-type stiffness maxima and the directions of common planes from this family differ by about 5 degrees, because the energy landscape is not symmetric around the minima.

From (i) – (iii) we conclude that the angular distribution of phase boundaries in the maze pattern is a subtle fingerprint of the underlying phase boundary anisotropy, even if skewed towards nearest-by corners of the Wulff shape. The observed skew indicates that phase boundaries move away from corners (and metastable orientations) towards the nearest-by energy minimum. Because of the skewed distribution a "reverse engineering" approach to construct energy landscapes from the analysis of maze patterns seems cumbersome. Nonetheless the quantitative analysis of early maze patterns allows concluding about the general features of the anisotropy at case.

*3.3 Maze-to-lamellar transition and associated defect evolution*

The phase boundary alignment process during the maze-to-lamellar transition entails the development of an associated pattern of defects which gradually settles into a network of fault lines. The defect evolution is clearly visible from the *Gabor filter* quality index displayed in fig. 5c. We further examine the defect structure in more detail and with reference to the location of corners in the Wulff shape computed from the 2D gamma-plot. Results are presented for a region of interest from section E3, basically a zoomed-in region near the most prominent nodular core structure of the maze. Fig. 8a shows the respective region with phase boundaries being outlined with an additional contour line. The contouring makes the defect lines appear more clearly as distinct lines in the pattern. Taking advantage of the known Wulff shape (fig. 8c) one can redraw the fault lines as dotted lines in colors which correspond to the location of the distinct corners, as shown in fig. 8b. It is obvious to see that the fault lines display directions which closely follow the corners of the Wulff shape and that characteristic intersection angles of e.g. 40°, 90°, 130° and 138° are present in the fault line pattern. This also holds for lines which connect physically present corners.

The result is not surprising, since phase boundaries aligned to corners or metastable orientations will disappear along with the evolution of lamellar order. Furthermore, the gradual alignment process cannot be accomplished without entailing defect formation, because the phase boundary energy minima for this eutectic and the given OR are separated from each other by corners. While being so obvious, the result is yet new: it brings a valid and novel aspect to the old dispute about the origin of fault lines in lamellar eutectics.

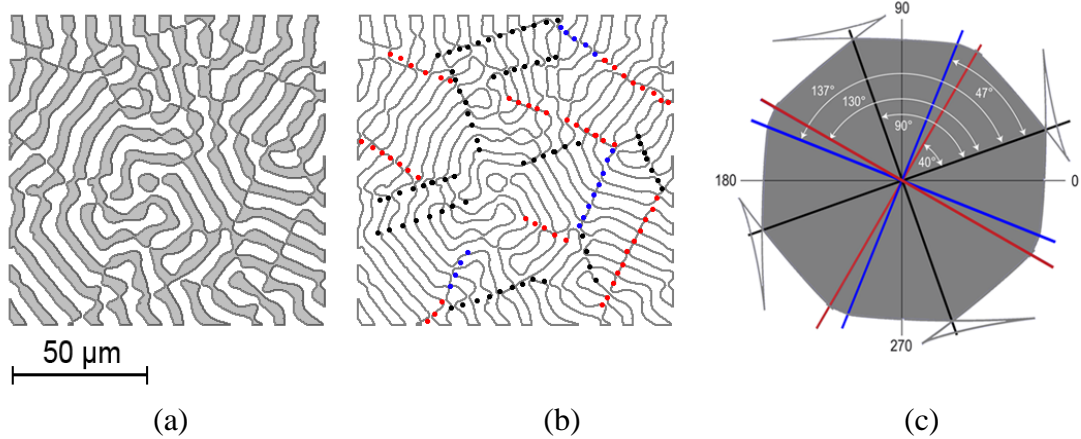

(a) (b) (c)

**Fig. 8:** The defect structures in the maze pattern **(a)** are fault lines alined to the corners of the Wulff shape **(b, c)** and display characteristic intersection angles of e.g. 40, 90 and 130° etc.

Fault lines are known to be the characteristic defects in a unidirectional solidified lamellar eutectic pattern [25, 26]. They were proposed to be growth inherent defects [27] necessary for allowing lamellar spacing adjustments during growth. They were proposed to result from a zig-zag instability of lamellar interfaces with an instability threshold of $0.85\lambda_m$, where $\lambda_m$ is the minimum-undercooling spacing [28]. They were proposed to be boundaries between crystallographically disoriented domains, e.g. subgrain boundaries [29, 30]. The results presented here show that, in the eutectic grain that we selected, fault lines were created during the maze-to-lamellar transition being the pattern's response to corners of the Wulff shape. One may further advance the conjecture that the fault line density of a eutectic grain will increase with the number of corners in the Wulff shape of the solid-solid phase boundary energy. In the absence of corners, e.g. for eutectics with small anisotropy of the phase boundary energy one could potentially obtain fault-free lamellar structures. The late stage of fault line accommodation in the lamellar pattern is not discussed here, but will be presented in a forthcoming paper.

## 4. Summary and outlook

We performed an experimental investigation of the crystallography, pattern and defect evolution during the maze-to-lamellar transition in a eutectic Al-Al$_2$Cu grain. The focus was placed on the early stages of directional solidification being probed in successive transverse section planes by serial sectioning. For the selected grain the crystal orientation and orientation relationship (OR) was determined by electron backscatter diffraction. The associated phase boundary energy landscape, e.g. the gamma-plot, was known from molecular dynamic computations [12]. It was thus possible to follow the phase boundary alignment process during the maze-to-lamellar transition and to discuss the experimental observations with reference to the phase boundary energy as function of the boundary location. For the selected grain the phase boundary energy landscape shows a pronounced anisotropy, with distinct energy minima separated by regions of high energy, i.e. corners (and forbidden orientations) in the Wulff shape. The phase boundaries were shown to align slowly to one of the energy minima, being selected amongst other. The following results and conclusions were obtained:

(i) during the maze-to-lamellar transition the Al-Al$_2$Cu phase boundaries slowly align to one distinct energy minimum, rather than remaining in the maze pattern. The exit from the maze is driven by a small crystallographic bias: the c-axis of the two crystals Al and Al$_2$Cu are not perfectly parallel, as in the ideal OR at case, but disoriented by about 2.5°. This disorientation breaks the symmetry of the OR favoring one over all other potential common planes associated with a low phase boundary energy. It was shown that phase boundaries gradually align to this specific plane, such that the normal direction to phase boundaries is parallel to the local maximum of phase boundary stiffness.

(ii) the maze-to-lamellar transition is slow and not fully finished after 5.5 mm of unidirectional growth. The phase boundary alignment is slowest in early stages of the maze evolution where distinct energy minima compete and accelerates once phase boundaries at competing locations have been outgrown. The lamellar pattern is not fully established after about 11 mm of growth and bears characteristic traces of the phase boundary energy landscape. While the overall transition is slow, local pattern changes can be very fast, especially upon disappearance of unfavorable alignments.

(iii) image analysis using the *Structure Tensor* has been performed to quantitatively determine the phase boundary orientation distribution with an angular resolution of 0.5°. The measurements have been used to create frequency plots in polar coordinates and to explore if the measured frequency distribution in the early maze is a direct fingerprint of the phase boundary energy landscape. The results show that it is not straightforward to map the equilibrium Wulff-plot (or gamma-plot) from the frequency distribution alone, mainly because frequency maxima are skewed towards nearest-by corners. It is however possible to first identify the corners of the Wulff-plot from the orientation of defect lines and then estimate the location and nature of energy minima from the frequency distribution and the given EBSD data.

(iv) the maze-to-lamellar transition entails the formation of defects which are fault lines. The fault lines display directions which closely follow the corners of the Wulff shape, leading to characteristic intersection angles between the fault lines in the defect network. The late stage of fault line accommodation in the final lamellar pattern will be presented in an upcoming publication.

Taken together the results showed how the solid-solid phase boundary anisotropy determines the pattern and defect evolution in the lamellar eutectic Al-Al$_2$Cu. Complementary experiments are currently conducted to further refine the analysis by replacing serial sectioning with X-ray tomography, such that more detailed information will be available regarding the local mechanisms of phase boundary motion and defect nucleation/propagation.

**Acknowledgements**

The authors gratefully acknowledge funding through the European M-Era.Net Project "ANPHASES" by DFG and ANR under grants number HE 6938/2-1 and ANR-14-MERA-0004, respectively. We would like to thank the project partners Mathis Plapp from the Ecole Polytechnique de Paris and Jürgen Horbach from the Heinrich Heine Universität Düsseldorf for fruitful discussions, valuable comments and suggestions.